\documentclass[aps,prl,twocolumn,showpacs,preprintnumbers,amsmath,amssymb,superscriptaddress,floatfix]{revtex4}%
\usepackage{epsfig}%
\usepackage{dcolumn}
\usepackage{bm}
\begin{document}%
\title{\Large\bf\boldmath 
Spin Gap in Heavy Fermion UBe$_{13}$ 
}%
\author{Vyacheslav G.~Storchak}
\email{mussr@triumf.ca}
\affiliation{National Research Centre ``Kurchatov Institute'',
 Kurchatov Sq.~1, Moscow 123182, Russia}
\author{Jess H.~Brewer}
\affiliation{Department of Physics and Astronomy,
 University of British Columbia,
 Vancouver, BC, Canada V6T 1Z1}
\author{Dmitry G.~Eshchenko}
\affiliation{Bruker BioSpin AG, Industriestrasse 26, 8117 F{\"a}llanden, Switzerland}
\author{Patrick W.~Mengyan}
\affiliation{Department of Physics, Texas Tech University, 
 Lubbock, Texas 79409-1051, USA}
\author{Oleg E.~Parfenov}
\affiliation{National Research Centre ``Kurchatov Institute'',
 Kurchatov Sq.~1, Moscow 123182, Russia}
\author{Pinder Dosanjh}
\affiliation{Department of Physics and Astronomy,
 University of British Columbia,
 Vancouver, BC, Canada V6T 1Z1}
\author{Zachary Fisk}
\affiliation{Department of Physics and Astronomy, University of California, 
Irvine, CA 92697, USA}
\author{James L.~Smith}
\affiliation{Materials Science and Technology Division, 
Los Alamos National Laboratory, Los Alamos, NM 87545, USA}
\vfil 
\date{22 April 2013} 
\begin{abstract}
Muon spin rotation ($\mu^+$SR) experiments 
have been carried out in the heavy fermion compound UBe$_{13}$ 
in the temperature range from 0.025~K to 300~K and 
in magnetic fields up to 7~T, supported by resistivity 
and magnetization measurements.  
The $\mu^+$SR 
spectra are characteristic of formation of 
a spin polaron which persists to the lowest 
temperature.  
A gap in the spin excitation spectrum of $f$-electrons 
opens up at about 180~K, 
consistent with anomalies discovered in resistivity, heat capacity, NMR 
and 
optical conductivity measurements of UBe$_{13}$.  
\end{abstract}
\vfil
\pacs{71.27.+a, 74.72 Kf, 74.20.Mn, 76.75.+i \\ }
\maketitle%
In strongly correlated metallic materials, 
the interplay between local 
spins and itinerant electrons 
determines the spin fluctuation spectrum.  
In a 3$d$-electron system, such a spectrum emerges as a result of 
a strong dominance of the Fermi energy over 
magnetic energy 
\cite{Moriya1985}.  
By contrast, in 
$f$-electron itinerant systems 
the Fermi energy is heavily renormalized down to 
the scale of the 
magnetic energy 
\cite{Coleman2006}, 
resulting in 
a strong influence of the spin dynamics 
and spectral weight of spin fluctuations 
on transport, magnetic and thermodynamic properties of a system.  
A depletion of spectral weight upon cooling 
may indicate the opening of a gap (or pseudogap) 
in the spectrum of spin fluctuations.  
 
The concept of a pseudogap has recently become 
essential 
for understanding strongly correlated electron systems 
\cite{Basov2005}.  
Particularly interesting is the opening of a gap 
in a system of 
$f$-electron spin excitations 
\cite{Wiebe2007}.  
At low temperature, such systems display a continuous transition 
from an array of uncorrelated local magnetic moments 
to a 
Fermi-liquid (FL) phase in which $f$ electrons 
are strongly hybridized with conduction band electrons 
\cite{Fisk1988}.  
As a result, 
$f$ electrons 
not only 
create a sharp resonance on the Fermi level, 
which gives rise to a large effective mass $m^*$ of quasiparticles, 
but also transfer their magnetic entropy to the Fermi surface.  
In this case a gap may open up 
as in URu$_2$Si$_2$ 
at $T=17.5$~K \cite{Wiebe2007}.  
In this Letter, we present experimental results that are 
explained 
by a spin gap opening in UBe$_{13}$ at $T$=180~K.  
 
UBe$_{13}$ crystallizes in the cubic NaZn$_{13}$ structure 
with lattice constant $a=1.025$~nm.  
The U atoms form a simple cubic sublattice 
with a rather large U-U spacing, 
suggesting 
strong hybridization with the itinerant carriers 
\cite{Pfleiderer2009}.  
Unlike in other uranium-based HF compounds, 
there is no evidence for static magnetic order in UBe$_{13}$.  
In UBe$_{13}$, superconductivity (SC) arises 
from a paramagnetic (PM) metallic phase 
as a cooperative phenomenon involving heavy quasiparticles 
that form pairs \cite{Ott1983,Pfleiderer2009}.  
At high temperature, the susceptibility $\chi$ exhibits 
a Curie-Weiss behavior with $\mu_{\rm eff}\approx 3.36~\mu_B$, 
which deviates from this 
law below 
200~K 
and levels off below 20~K \cite{Brison1989}.  
 
Interest in UBe$_{13}$ is reinforced by various observations of 
non-Fermi-liquid (NFL) behavior 
\cite{Brison1989,Oeschler2003,Gegenwart2004,Flouquet2005,Pfleiderer2009}, 
often explained by the proximity 
of 
quantum critical point 
\cite{Oeschler2003,Gegenwart2004,Stewart2001}, 
which challenges 
the validity of the quasiparticle concept.  
However, 
various experiments indicate a dominant role of quasiparticles, 
although the specific quantum states that might replace the 
Fermi liquid 
remains unclear \cite{Stewart2001}.  
 
Quite remarkably, FL behavior 
is restored by application 
of a strong magnetic field 
\cite{Gegenwart2004}.  
Another remarkable feature which requires explanation is that 
while at low temperature UBe$_{13}$ 
displays 
coherent quasiparticle behavior, 
at higher temperature 
it reveals 
an {\sl incoherent\/} metallic state 
dominated by spin-flip scattering 
\cite{Gegenwart2004,Brown2003}.  
Both NFL and loss of coherence phenomena 
cannot be explained by quadrupolar Kondo effect \cite{Oeschler2003}.  
Furthermore, 
energy band calculations 
produce an $m^*$ value 
at least an order of magnitude too low \cite{Liu1997}.  
Moreover, the band theory fails to explain the loss of coherence.  
 
\begin{figure}[t] 
\begin{center} 
\includegraphics[width=\columnwidth,angle=0]{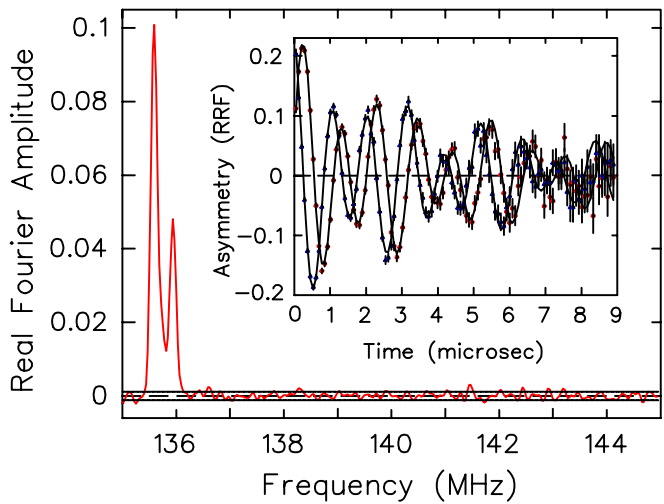} 
\vspace*{-6mm} 
\caption{ 
 Frequency spectrum of muon spin precession in UBe$_{13}$ 
 in a transverse magnetic field $B=1$~T at $T=150$~K.  
 Inset: same spectrum in the time domain in a rotating reference frame.  
 The two-frequency precession characteristic of a muon-electron 
 hyperfine coupled state is evident in both domains.  
} 
\end{center} 
\vspace*{-10mm} 
\label{FFT150K} 
\end{figure} 
 
In order to identify a fundamental electronic state 
consistent with all the crucial experimental results, 
different polaron models have been proposed 
\cite{Liu1997,Kasuya2000}.  
Although both {\sl electronic\/} and {\sl magnetic\/} 
polaron models may account for 
NFL behavior and the loss of coherence 
and produce a correct $m^*$ \cite{Liu1997}, 
the former fails to explain the strong influence of a magnetic field.  
By contrast, a spin polaron (SP) model \cite{Kasuya2000} 
claims to establish an electronic state which may help to reconcile 
theoretical treatment with the experiment.  
In particular, the SP model \cite{Majumdar1998} offers 
a straightforward explanation of a major puzzle 
of UBe$_{13}$ --- a remarkable magnetoresistivity (MR) 
\cite{Brison1989} --- which can hardly find explanation 
within any other model.  
Both MR and magnetostriction indicate that the carrier number 
is a strong function of magnetic field 
as if {\sl ``the carrier is released by $B$''} \cite{Flouquet2005}.  
The remarkable sensitivity of the electron transport 
to the magnetization 
(an order of magnitude stronger than in hole-doped manganites) 
shows that carrier localization into SP, 
and its release 
by a magnetic field, 
may be a missing 
key ingredient of HF models.  
A SP model 
\cite{Majumdar1998} 
predicts 
that the narrowness of the conduction band may cause 
the carriers to self-trap as SP.  
Recent observation of SP in 
magnetoresistive Lu$_2$V$_2$O$_7$ 
\cite{Storchak2013} supports this model.  
In this Letter, we present observation of SP in UBe$_{13}$. 
It is this observation which allows detection of a spin gap.  
 
Single crystals for the current studies display lattice constants, 
resistivity behavior, $T_c$ and effective magnetic moments 
consistent with literature data.  
Time-differential $\mu^+$SR experiments \cite{Brewer1994} 
using 100\% spin-polarized positive muons 
implanted into these samples 
were carried out on the M15 muon channel at TRIUMF 
using the {\it HiTime\/} and {\it DR\/} spectrometers.  
At high temperature, in magnetic fields {\boldmath $\vec{B}$} transverse 
to the initial muon polarization, 
Fourier transforms of the $\mu^+$SR time spectra exhibit 
a single line at the 
muon frequency $\nu_\mu = \gamma_\mu B/2\pi$ 
(where $\gamma_\mu = 2\pi \times 135.53879$~MHz/T 
is the muon magnetogyric ratio), 
which coincides with that detected 
in a CaCO$_3$ reference sample used for independent measurements of $B$.  
However, below $T^* \approx 180$~K such simple spectra change abruptly 
to reveal a characteristic doublet (Fig.~1) 
which persists down to the lowest temperature (Fig.~2).  
 
\begin{figure}[t] 
\begin{center} 
\includegraphics[width=0.9\columnwidth,angle=0]{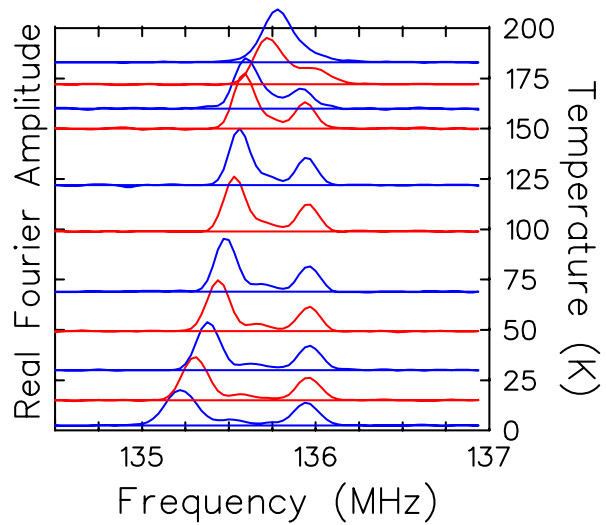} 
\vspace*{-3mm} 
\caption{ 
 Fourier transforms of the muon spin precession signal in UBe$_{13}$ 
 in a transverse magnetic field of 1~T 
 at different temperatures.  
 Characteristic SP doublet persists 
 down to the lowest measured temperature T=0.025~K.  
} 
\end{center} 
\vspace*{-8mm} 
\label{UBe13-evolution} 
\end{figure} 
 
Previous observations of such two-line spectra 
\cite{Sonier2000}
prompted a suggestion of two 
inequivalent sites occupied 
by positive muons in UBe$_{13}$.  
However, the two lines do not follow the temperature dependence of 
the magnetization, which indicates that the muon does not stay bare 
and does not act as a local magnetometer.  
Instead, while one line goes up in frequency 
the other goes down as temperature varies --- 
a signature of a muon-electron bound state \cite{Brewer1994}.  
Similarly, if the muon stayed bare, 
both lines' frequencies should stay constant 
as the susceptibility levels off below 20~K, 
which contradicts the experiment.  
Furthermore, a two-site assignment suggests 
a temperature independent strict 1:2 
muon site occupation ratio \cite{Sonier2000} 
while experiment shows a temperature dependent 
amplitude ratio well below 0.5.  
Finally, the departure from linear magnetic field dependences 
of the two frequencies allows one to rule out 
possible Knight shifts within the bare muon scenario.  

Instead, the doublet on Fig.~1 and Fig.~2 is a fingerprint of 
a coupled $\mu^+ e^-$ spin system in high transverse field 
\cite{Brewer1994,Storchak2009,Storchak2009a,Storchak2010, 
Storchak2011,Storchak2012,Storchak2013}: 
the two lines correspond to two muon spin-flip transitions 
between states with electron spin orientation fixed, 
the splitting between them being determined by 
the muon-electron hyperfine interaction $A$ \cite{Storchak2009}.  
In a magnetic system, 
an electron's energy depends strongly on the magnetization, 
with the minimum energy being achieved by FM ordering \cite{Nagaev2002}.  
Then the strong exchange interaction $J$ 
between a carrier and local $f(d)$ moments 
can cause electron localization into a ferromagnetic (FM) ``droplet'' 
over the extent of its wavefunction (typically, the first coordination sphere) 
in a PM (or AF) sea \cite{deGennes1960}.  
This charge carrier, accompanied by reorientations 
of local spins to form its immediate FM environment, 
together behave as a single quasiparticle with a giant spin $S$ --- 
a {\sl spin polaron\/} \cite{deGennes1960,Nagaev2002}.  
In the process of electron localization into a SP in a metal 
the exchange energy gain upon transition from the PM to the FM state 
is opposed by the increase 
of the electron kinetic energy 
(the entropy term due to ordering within the SP 
becomes significant only at very high $T$) 
\cite{Storchak2009,Storchak2010,Storchak2011}.  
 
High magnetic field destroys the SP because 
in high $B$ the spins are already polarized, 
so that the exchange coupling of the carrier with these spins 
offers no energy advantage to compensate 
the increase in kinetic energy due to localization.  
Application of a magnetic field thus releases 
the carrier from SP into the 
conduction band --- 
a process which offers not only an 
explanation of the huge 
negative MR but also reveals the reason why 
the carrier number is a strong function of 
magnetic field in UBe$_{13}$.  
 
Such SP states with the electron confined in a 
$R \approx 0.2-0.5$~nm FM ``droplet'' 
attached to a positive muon 
are found in strongly correlated insulators 
\cite{Storchak2009a,Storchak2012a}, 
semiconductors 
\cite{Storchak2009,Storchak2009a,Storchak2012} 
and metals 
\cite{Storchak2010,Storchak2011}.  
In metals, an 
itinerant SP is captured by the muon to 
exhibit the characteristic $\mu^+e^-$ hyperfine splitting 
\cite{Storchak2010,Storchak2011} 
through the frequency splitting $\Delta\nu$ 
between two SP lines (Fig.~3).  
Within a mean field approximation, $\Delta\nu$ 
follows a Brillouin function 
\cite{Storchak2009,Storchak2010,Storchak2011}.  
For $g \, \mu_{_{\rm B}} B \ll k_{_{\rm B}} T$, 
\begin{equation} 
 \Delta\nu = A \left[\frac{g \, \mu_{_{\rm B}} B}{3k_{_{\rm B}} (T-\Theta)}\right] 
  ({\cal S}+1) \; .  
\label{eq:Splitting} 
\end{equation} 
A strong deviation from the Curie-Weiss law, 
which lies in the heart 
of HF behavior, restricts UBe$_{13}$ 
to the small $B/T$ limit 
so that Eq.~(\ref{eq:Splitting}) stays valid 
in the entire measured $B$ range, 
in contrast to other systems which support SP 
\cite{Storchak2009,Storchak2010,Storchak2011}.  
Fitting $\Delta\nu(B)$ for $T$ between 15~K and 150~K with $\Theta$ 
found from magnetization measurements (see inset in Fig.~4) 
and taking into account the dominance of the orbital magnetic moment 
which causes $g=0.8$ \cite{Yaresko2003} 
gives $A=45(5)$~MHz and $S=8.5(0.5)$.  
From the value of $A \propto R^{-3}$ we get $R = 0.25(1)$~nm, 
which rather remarkably indicates a maximum overlap 
of corresponding $s$ and $f$ wavefunctions 
within the SP \cite{Storchak2009,Storchak2010,Storchak2011}, 
as the U-U distance is 0.5124~nm 
and the radius of the $f$-orbital is 0.0527~nm.  
This is consistent with the muon sitting 
in between the two U ions \cite{Sonier2000} 
that captured a SP consisting of an electron 
whose wavefunction overlaps $f$-orbitals of 
said 
U ions, each having a magnetic moment 
$\mu_U=g\mu_{\rm B}(S+1/2)/2=3.6\mu_{\rm B}$, 
which is close to the $\mu_{\rm eff}$ found from susceptibility measurements.  
 
\begin{figure}[t] 
\begin{center} 
\includegraphics[width=\columnwidth,angle=0]{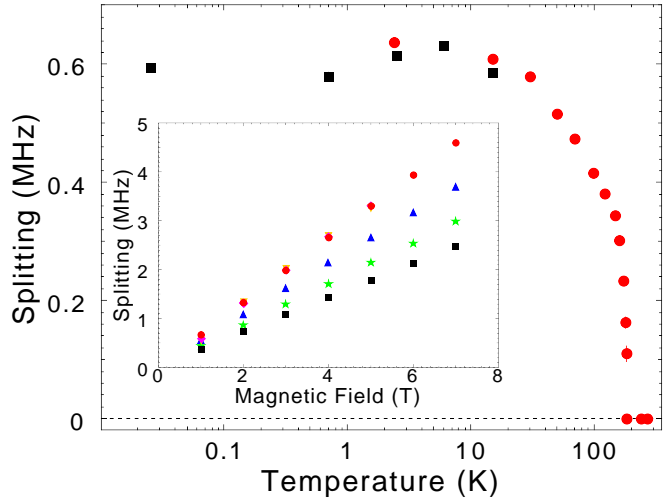} 
\vspace*{-7mm} 
\caption{ 
 Temperature dependence of the SP frequency splitting $\Delta\nu$ 
 in UBe$_{13}$ in a magnetic field of $B=1$~T measured using HiTime (circles)
 and DR (squares) spectrometers.  
 Inset: magnetic field dependences of $\Delta\nu$ at $T=150$~K (squares), 
 $T=99$~K (stars), $T=50$~K (triangles), $T=15$~K (circles), $T=0.7$~K (nablas) 
 and $T=0.025$~K (diamonds).  
} 
\end{center} 
\vspace*{-10mm} 
\label{UBe13-split} 
\end{figure} 
 
In a PM, strong spin exchange with the environment 
\cite{Brewer1994} 
would result in rapid spin fluctuations of the SP electron, 
averaging the hyperfine interaction to zero, 
which in turn would result in a collapse 
of the doublet into a single line at $\nu_{\mu}$ 
(see \cite{Storchak2011a} for details) 
unless the SP spin $S$ is decoupled from the local spins 
\cite{Storchak2009,Storchak2010,Storchak2011a}.  
Such decoupling is possible in high $B$ 
when the Zeeman energy of $S$ 
exceeds an exchange interaction $J$ between local spins --- 
this is the case in magnetic 
insulators 
where the SP doublet is detected up to very high temperature 
\cite{Storchak2009,Storchak2009a,Storchak2013}.  
In metals, RKKY interactions make 
$J$ much stronger, so that 
decoupling would require $B$ strengths 
inaccessible in the current experiment.  
That's why we see a single line above 180~K, 
which by no means 
indicates the absence of the SP above 180~K --- 
just active spin exchange with the environment 
\cite{Storchak2009,Storchak2010,Storchak2011a}.  
The abrupt appearance of a SP doublet at 180~K (Fig.~3) 
we interpret as a result of opening of a spin gap 
that eliminates low-lying spin excitations, 
making spin exchange of $S$ with its environment ineffective.  
This explanation is consistent with anomalies in specific heat 
\cite{Felten1986}, 
NMR \cite{Clark1988} 
and optical conductivity 
\cite{Brown2003}, 
which are discussed in terms of crystal field splitting of 
the 5$f^3$ 
U ion 
by a characteristic energy of $\sim 180$~K.  
 
Our measurements of the electric resistivity $\rho$ 
of the same single crystals (Fig.~4) 
confirm a spin gap opening in UBe$_{13}$ at about 180~K.  
A basic behavior of resistivity lies in the context of 
carrier scattering in metals with local spins 
\cite{Kasuya1956}, 
as in UBe$_{13}$ below 300~K 
$\rho$ is dominated by carrier scattering on spins 
\cite{Andraka1991}.  
At high temperature, 
it consists of a temperature-independent local scattering term 
and a $1/T$ term due to scattering on paramagnons \cite{Kasuya1956}.  
At lower temperature down to $\Theta$, Kondo scattering takes over.  
A crossover from $\rho(T)=A/T+C$ 
to $\rho(T) = D \ln(E/T)$ 
at about 180~K signifies a characteristic change 
in the spin fluctuation spectrum of the system.  
(Fitting the data gives the following values: 
$A = 8.5 \times 10^{-3}$~Ohm$ \cdot $cm$ \cdot $K, 
$C = 94 \times 10^{-6}$~Ohm$ \cdot $cm, 
$D = 47.4 \times 10^{-6}$~Ohm$ \cdot $cm 
and $E=3500$~K.) 
At high temperature, the largest contribution to $\rho$ 
comes from carriers with small momentum $q\sim 0$ 
\cite{Kasuya1956} 
scattered by low-lying excitations 
dwarfing both $1/T$ and $C$ 
as the spin gap opens up, eliminating such excitations.  
On the other hand, 
an opening of the spin gap 
promotes resonant Kondo scattering 
effective at a significant $q$ and energy 
\cite{Abrikosov1988}.  
Moreover, disappearance of the low-lying spin excitations 
causes strong suppression of long-wavelength magnetic fluctuations 
accompanied by deviation of $\chi$ from the Curie-Weiss law 
at 180~K seen in $\Theta(T)$ dependence (inset in Fig.~4).  
An opening of the spin gap might be explained by 
the position of an $f$-level 
lying 180~K below the Fermi surface.  
 
\begin{figure}[t] 
\begin{center} 
\hspace*{-4mm} 
\includegraphics[width=\columnwidth,angle=0]{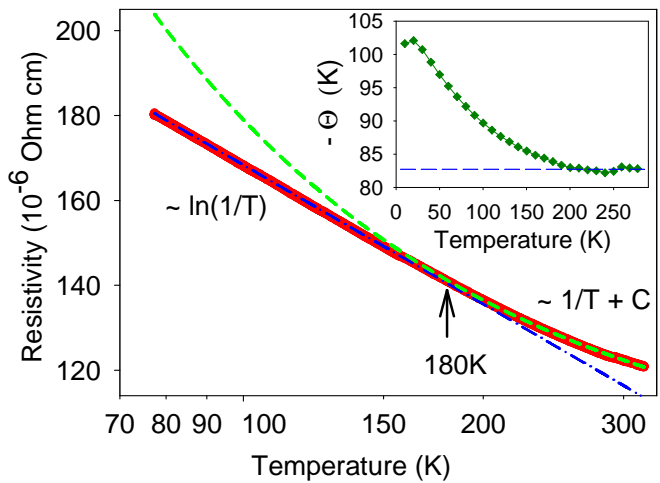} 
\vspace*{-2mm} 
\caption{ 
 Temperature dependence of the electrical resistivity of UBe$_{13}$ 
 (red: experimental points; green and blue dotted lines: 
 fitting to corresponding functions).  
 Inset: $\Theta(T)$ dependence extracted from 
 magnetization measurements.  
} 
\end{center} 
\vspace*{-8mm} 
\label{UBe13-Resist} 
\end{figure} 
 
A standard approach to HF systems 
considers as a starting point 
a set of strongly localized $f$-electrons; 
the appearance of a new energy scale 
results from hybridization of conduction electrons 
with local $f$ moments so that 
heavy 
quasiparticles appear on the Fermi surface.  
An alternative approach \cite{Fulde1995} 
starts from a delocalized band carrier 
whose transport depends upon the strength of its coupling with 
excitations of the medium: in the limit of strong coupling 
an electron accompanied by lattice or spin excitations 
forms a quasiparticle --- a polaron.  
As is the case for the 
well-known lattice polaron, 
formation of a spin polaron may profoundly renormalize 
the bare electron band (bandwidth $\Delta_0 \sim 1$~eV) into an 
extremely narrow ($\Delta_{\rm SP} \sim 10^{-4}-10^{-3}$~eV) 
SP band \cite{Nagaev2002}.  
At low temperature, 
such an SP band supports coherent SP dynamics \cite{Storchak2012}.  
As SP have spin higher than 1/2, 
they do not need to follow a FL state --- 
hence NFL behavior is possible.  
At still lower temperature, formation of spin bipolarons might cause SC 
\cite{Mott1993,Storchak2010}.  
Here, the opening of a spin gap is a ``must have" ingredient, 
as it protects paired electrons from 
spin exchange with the environment, 
which destroys pairs.  
At higher $T$, however, the SP dynamics occur 
on a background of strong coupling to spin fluctuations, 
which destroys coherence.  
A dramatic renormalization of the SP band 
is expected to go hand-in-hand with a 
significant increase of $m^*$, 
which may allow application of 
general concepts developed for 
coherent-to-incoherent crossover of 
the tunneling dynamics of heavy particles: 
suppression of coherence in a metal 
is expected at $T \sim \Delta_{\rm SP}$ \cite{Storchak1998}.  

In summary, a spin gap opens up in UBe$_{13}$ at 180~K, 
detected by spin polarons.  
Formation of the SP band may explain several 
mysteries of UBe$_{13}$: 
NFL behavior, loss of coherence in a metal, and huge magnetoresistance.  
Emergence of such SP bands might be a general phenomenon in HF systems 
\cite{CeCoIn5}.  
Within this picture, the SP is itself the celebrated heavy fermion.  
 
This work was supported by 
the NBIC Center of the Kurchatov Institute, 
NSERC of Canada 
and the U.S. DOE, Basic Energy Sciences (Grant DE-SC0001769).  
\vspace*{-4mm} 
 

\end{document}